# Comparison Clustering using Cosine and Fuzzy set based Similarity Measures of Text Documents

Manan Mohan Goyal[1], Neha Agrawal[2], Manoj Kumar Sarma[3], Nayan Jyoti Kalita[4]
*Department of CSE, Royal School of Engineering and Technology*
*Guwahati-35, Assam, India*
{[1]manangoyal3, [2]agrawal.n001, [4]nayan.jk.123}@gmail.com
[3]manojksarma@yahoo.com

***Abstract*— Keeping in consideration the high demand for clustering, this paper focuses on understanding and implementing K-means clustering using two different similarity measures. We have tried to cluster the documents using two different measures rather than clustering it with Euclidean distance. Also a comparison is drawn based on accuracy of clustering between fuzzy and cosine similarity measure. The start time and end time parameters for formation of clusters are used in deciding optimum similarity measure.**

## I. Introduction

Clustering is one of the widely used data mining techniques. It has various applications in classification, visualisation, document organization, indexing and collaborative filtering [1]. It is an automatic grouping of text documents so that documents in same cluster or group are more similar to each other and more dissimilar to documents in other groups. As we know that text data is growing rapidly thus clustering can help in organizing text collection for efficient and effective browsing.[2] This rapid rise can help in making clustering a highly active research area. With the dynamic nature of the data the algorithm should be able to deal with the new data that is constantly added to the database.

Text mining is a burgeoning new technology for discovering knowledge. Text mining extended the data mining approach to textual data and is responsible for finding useful and interesting patterns, models, directions or rules from unstructured texts. Clustering is referred to as unsupervised learning the data which is to be trained is not labelled and the main objective turns out to be find natural clusters among the given patterns. Clustering is grouping of abstract or physical objects into classes of similar objects. Initially the data is collected by selecting the sources to exploit and then the individual texts are retrieved from the selected sources. Data warehousing includes information extraction which is the process of analysing unrestricted texts to fetch the pre-defined information like entities, relationships and types of event. Data Exploitation includes Data Mining and Data Presentation [3].

The analysis has been conducted on 1000 text documents and the result has been recorded for both the similarity measures. The parameters start time and end time are used to calculate the total time taken by both the measures to cluster the documents.

## II. Literature Survey

Earlier, K-means algorithm has been used with Euclidean distance to cluster the documents [4]. As we know K-means is associated with the formation of centroids which is the term from Euclidean Geometry itself. Since childhood one has learn to classify between dogs and cats or apples and oranges but to let computer do this task is the tedious job. Computer has to take various measures for clustering even small amount of text [5]. The algorithm used should be scalable and accurate. Accuracy should determine large intra-cluster similarity and less inter-cluster similarity. Cluster can be of any shape and size and clustering algorithm should be able to identify cluster of any arbitrary shape. Some of the following tasks need to be done which takes input as a plain text document and output a set of tokens to be included in the vector model.

1) *Filtering:* The process of removing punctuation and special characters which do not have any value under vector models.
2) *Tokenization:* Sentences are split into individual tokens, typically word.
3) *Stemming:* The process of reducing words to their root words in order to improve accuracy and reduce redundancy.
4) *Stopword Removal:* The word which does not convey any meaning as a dimension in a vector space. Non-informative words are called as stopwords which includes prepositions, articles, conjunctions and certain high frequency words.
5) *Pruning:* Removes the word with very low frequency.
6) *TF-IDF:* The weight of each document is characterized by term frequency-inverse document frequency. The term frequency-inverse term frequency is selected by having the knowledge of domain. The weight of the term will be zero if the term does not occur in the document, and the weight will be large if it occurs more frequently. The importance of the term is captured by the term frequency.

$$Weight_{ij} = log(tf_{ij}) * log(N/N_i)$$

where,
N = documents in a collection,
Tf-idf = weight,
$tf_{ij}$ = Number of times term i occurs in document j,
$N_i$ = Number of documents, out of N, in which term i occurs.

## III. Related Work

In [6] clustering related to fuzzy sets are described. Fuzzy sets can be used can be used to compare different types of objects such as images. This paper describes the properties of fuzzy sets.

In [7] the basic concept of clustering is explained. It describes the various measures needed to cluster the text documents. It explains the vector space model and the unsupervised phenomena.

In [8] various clustering approaches and algorithms have been projected and been used for Document Clustering. This paper deals with the projection of various algorithms but still there are some issues which are not even overcome by this algorithms even. So there is a scope that in future that the combination of these algorithms will be used to overcome document clustering problems.

In [9] distance based similarity measures of Fuzzy sets are considered which have a high importance in reasoning methods handling sparse fuzzy rule bases. The rule antecedents of the sparse fuzzy rule bases are not fully covering the input universe.Therefore the applied similarity measure has to be able to distinguish the similarity of non overlapping fuzzy sets, too.

In [10] a detailed survey of the problem of text clustering is done. They have discussed the key challenges of the clustering problem, as it applies to the text domain. They have also discussed the key methods used for text clustering, and their relative advantages. A good clustering of text requires effective feature selection and a proper choice of the algorithm for the task at hand.

In [11] they have analysed Document Clustering on various datasets. By this analysis we can understand various conditions responsible for various Clustering used. The analysis also shows that the method works efficiently for large data. They have used unstructured datasets. The lower frequency terms are removed which improves clustering effectiveness and reduces the computational complexity.

## IV. Methodology

Our motive was to compare the total time taken by both the similarity measures and to prove that fuzzy take comparatively less amount of time as cosine and form better clusters. The steps of implementation are discussed as follows.

### A. *Accessing the Folder*

The data to be clustered is needed to be fetched from the desired directory or folder in order to apply the clustering algorithm. The top-level window with a title and border is called frame. These frames are used for accessing the folder where our text file resides. Documents are represented into vector space model which is used for information filtering, information retrieval, indexing and relevancy rankings. The formation of vectors of documents are required so that sparse matrix can be generated which will help in clustering.

### B. *Selection of centroids*

K-means algorithm is entirely based on the selection of centroids. Centroid is the mean value of all the parameters values in the cluster. The k-means algorithm is used and the similarity measure is taken an0sd accordingly the clusters are formed. Our algorithm used is repetitive and the algorithm continues until there are no further changes in our cluster.

### C. *Removal of Stopwords*

Stopwords are the words which are filtered out before or after the processing of the text documents. We have collected around 119 stopwords in our project and stored them in data folder. Every document is compared with these set of stopwords and these words are removed from the documents.

### D. *Removal of Stopwords*

Stemming is the term used to reduce the inflected word to its root or stem word. We have around 67 stemmed patterns in our project. Text documents are compared with these patterns and are stemmed.

*NOTE:*It should be noted that stopword removal should be applied before stemming so that the word does not get stemmed and remain in the document. E.g. If we take a word was and perform stemming operation before than stopword removal, then the word wa will be left in the document after removal of s. Now this word wa is neither a stopword or a stemword thus increasing the frequency of the document.

### E. *Similarity*

The Document vectors which are obtained need to be normalised to provide uniqueness and better calculation. We have use two similarity measure and calculated the time taken by both of them.

1) COSINE:

$$\mathrm{sim}(d_1, d_2) = \frac{\vec{V}(d_1) \cdot \vec{V}(d_2)}{|\vec{V}(d_1)||\vec{V}(d_2)|}$$

2) FUZZY:

$$sim(C_1, C_2)_{\text{fuzzy}} = \frac{|F_{C_1} \cap F_{C_2}|}{|F_{C_1} \cup F_{C_2}|}$$

3) NORMALISATION:

similarity = Math.sqrt(similarity);
similarity = 1.0 / similarity;

Normalisation is used to normalize the vector. If we have to normalize the vector around x axis then we divide it by its square root.

### *F. Generation of output*

The output of the following code forms the cluster and gives the total time taken to form these clusters. The output for the time taken is obtained in the text file formed in our desktop window. Clusters are formed in the folders where our text documents reside.

## V. EXPERIMENTATION AND RESULTS

Based on the results obtained from the two similarity measures, it has been found out that the output obtained from the fuzzy similarity measures is better (time factor). We have run our codes in command prompt.

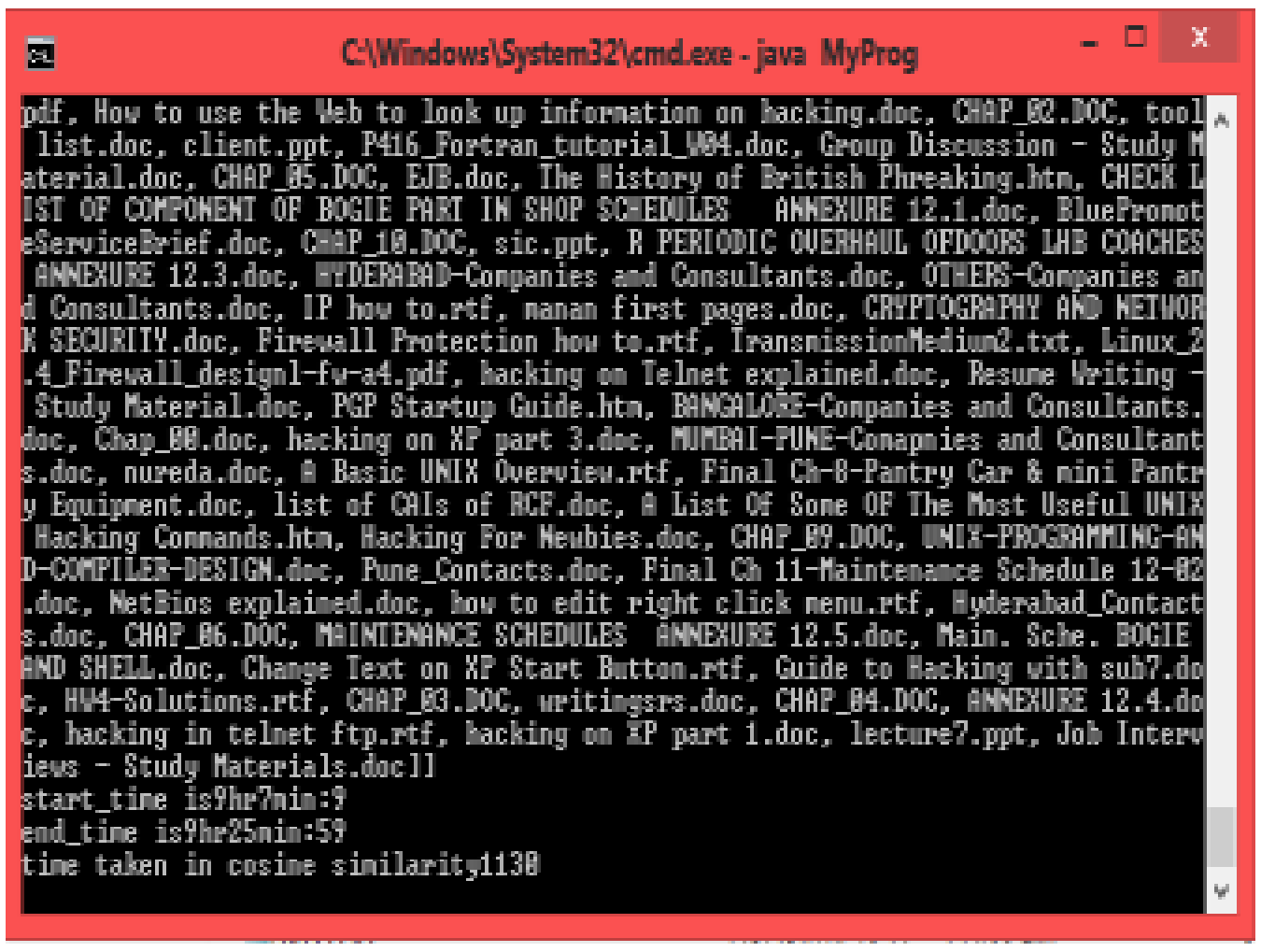

Fig. 1 Output for Cosine Measure

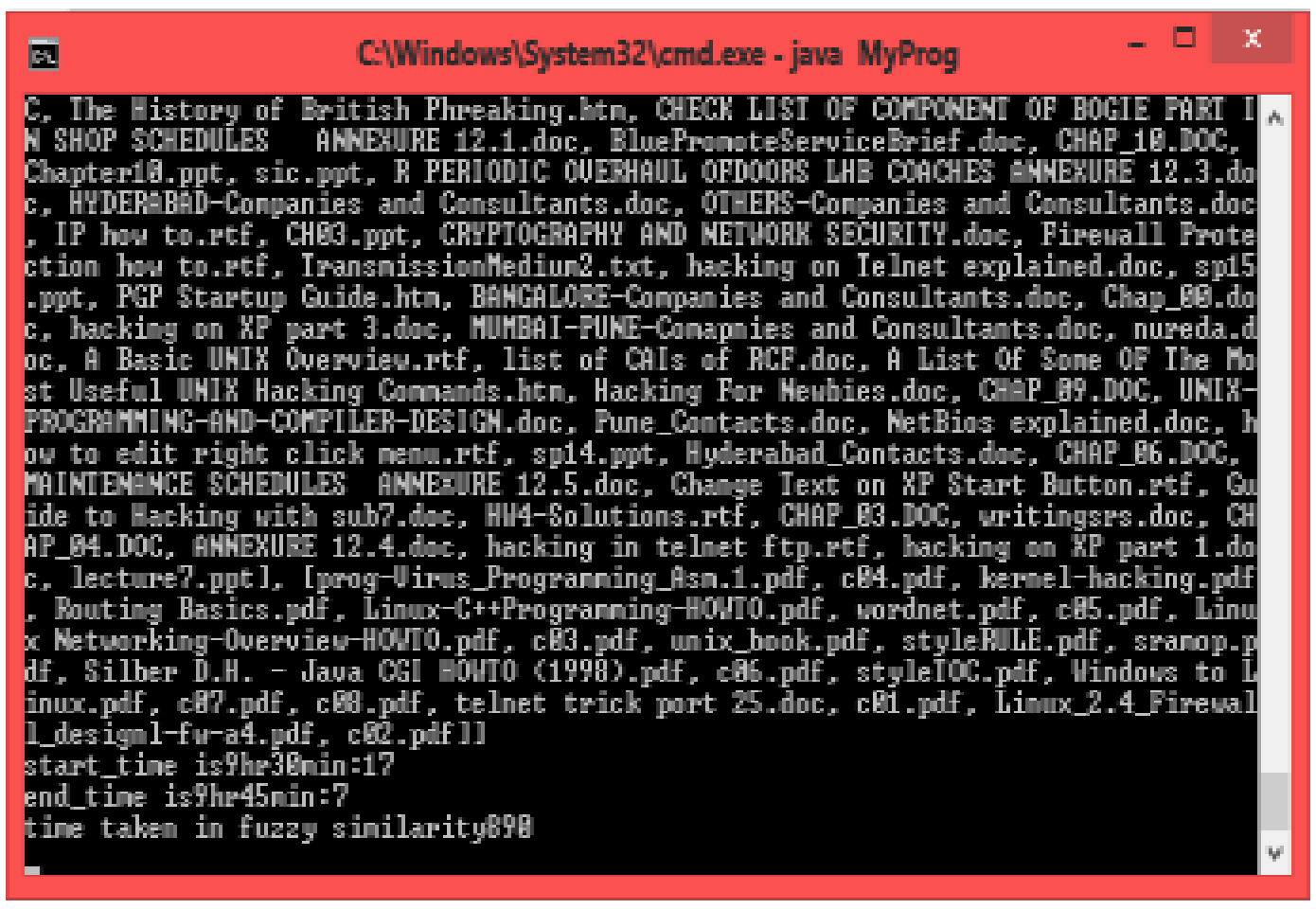

Fig. 2 Output for Fuzzy similarity Measure

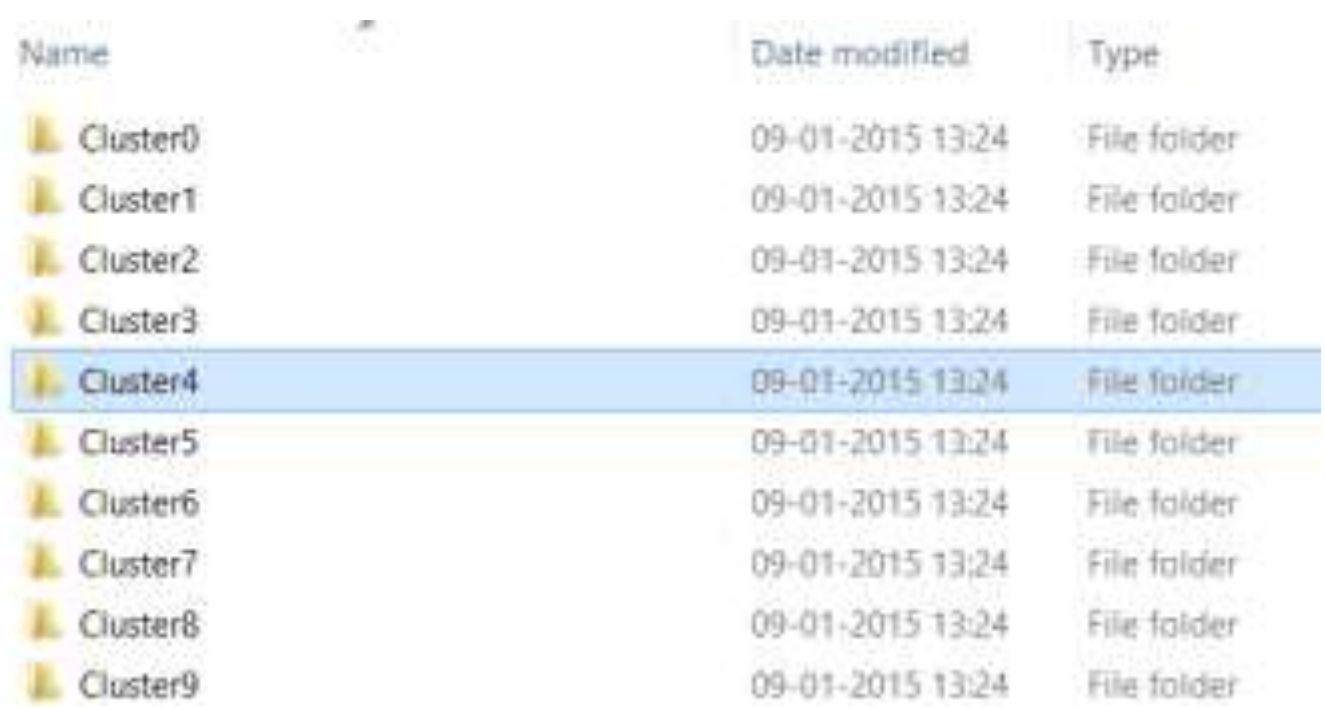

Fig. 3 Formation of Clusters

| Time | Cosine Measure | Fuzzy |
|---|---|---|
| **Start Time (System Time)** | **9:07:09 PM** | **9:30:17 PM** |
| **End Time (System Time)** | **9: 25:59 PM** | **9:45:7 PM** |
| **Total Time** | **1130sec** | **890sec** |

Fig. 4 Comparison Table of result

The time obtained in cosine is comparatively more than fuzzy. Hence proving fuzzy to be better similarity measure with k-means algorithm clustering.

## VI. CONCLUSION

In this paper, we have tried to compare the two similarity measures both cosine and fuzzy similarity measure using the k-means algorithm. Our objective was to produce the efficient and effective clusters after the application of the algorithms. The total time calculated turned out to be less in fuzzy similarity measure as compared to cosine similarity, hence proving fuzzy to be the better similarity measure. The real world is dynamic in nature with the database being constantly updated. It is not possible to run the clustering algorithm to every new database being added to it. The clustering algorithm must be incremental so that the whole databases need not to be used every time for clustering.